\documentclass[reprint,prb, showpacs]{revtex4-1}% Physical Review B

\usepackage{graphicx}
\begin{document}

\preprint{Draft-0106}

\title{Effects of Anisotropic Charge on Transverse Optical Phonons in NiO}

\author{H. Uchiyama$^{1,2}$, S. Tsutsui$^{2}$, A.~Q.~R Baron$^{1,2}$}
\affiliation{
1 Material Dynamics Laboratory, RIKEN, Sayo, Hyogo, 679-5148, Japan.\\
2 Japan Synchrotron Radiation Research Institute, Koto, Sayo, Hyogo, 679-5198, Japan.
}%

\date{\today}
\begin{abstract}
 Phonon dispersion of detwinned NiO is measured using inelastic x-ray scattering. It is found that, near the zone center,  the energy of the  transverse optical phonon mode polarized parallel to the antiferromagnetic order is $\sim$1 meV lower than that of the mode polarized perpendicular to the order,  at room temperature.  This is explained via anisotropic polarization of the Ni and O atoms, as confirmed using a Berry's phase approach with first-principles calculations.  Our explanation avoids an apparent contradiction in previous discussions focusing on Heisenberg interaction.
\end{abstract}

\pacs{75.50.Ee, 75.30.Gw, 77.84.Bw, 63.20.dd}
\maketitle

%J1 negative for NiO
Transition-metal mono-oxides are fundamental materials for studying properties of magnetic and strongly correlated systems.
Among them, NiO and MnO are antiferromagnetic (AFM) insulators with similar crystal structure and physical properties. The AFM order and lattice contraction occur along the [111] direction below the N\'{e}el temperature ($T_{\rm N}$=523 K for NiO). The order consists of ferromagnetic (111) planes. Above $T_{\rm N}$, both have a rock-salt structure. NiO and MnO are relatively simple, because they have non-degenerate electronic ground states, and are free from Jahn-Teller effects, unlike FeO and CoO.\cite{Kanamori} 

The magnetism of these materials is often discussed in terms of a superexchange mechanism. MnO has antiferromagnetic interactions both in the nearest neighbor ($J_1$) and next nearest neighbor ($J_2$) exchanges, and these interactions reproduce the experimental results.\cite{ExStriction,J_phonon} On the other hand, $J_1$ for NiO still remains uncertain, perhaps because it is much smaller than the antiferromagnetic superexchange $J_2$ interaction, which is responsible for the AFM order. For example, local spin density approximation (LSDA)+$U$ calculations give ferromagnetic $J_1$,\cite{J_phonon} while some calculations such as GW show antiferromagnetic $J_1$.\cite{GW} Experimentally, a measurement of spin-wave dispersion indicates ferromagnetic $J_1$,\cite{SpinWave} while one of magnetic susceptibility suggests antiferromagnetic $J_1$.\cite{Magnetic} 

Another issue is that LSDA+$U$ calculation results are not easily reconciled with the exchange interaction picture of NiO. According to Refs.~\cite{ExStriction, Yamada}, the lattice distortion is dominated by $J_1$ with no contribution from $J_2$, and antiferromagnetic $J_1$ causes the contraction in [111], if $|J_1|$ decreases with increasing the distance between the nearest-neighbor Ni atoms. Based on this discussion, the calculated ferromagnetic $J_1$\cite{J_phonon} is not consistent with the calculated\cite{Forceconst} (and observed) contraction in NiO, suggesting an additional ingredient is needed to understand the calculation results. 

The energy of transverse optical (TO) phonons can be used as a direct probe of microscopic coupling \cite{Massida}; TO modes that would be degenerate in a cubic rock-salt structure, are split at the zone center under the AFM order, with the energy of the mode polarized along the order ($E_{\parallel}^{\rm TO}$) different than that of the mode polarized in the plane perpendicular to the order ($E_{\perp}^{\rm TO}$). As discussed in Ref.~\cite{J_phonon} the sign of this difference ($E_{\parallel}^{\rm TO}$$-$$E_{\perp}^{\rm TO}$) can be linked to the sign of $J_1$. The picture is generally consistent for MnO, where the observed shift\cite{Chung,MnOIR} is in agreement with theory,\cite{J_phonon,Forceconst} however question remains about NiO, where preliminary measurements \cite{Chung} disagree with the LSDA+$U$ calculations. \cite{J_phonon, Forceconst}

%what is achieved 
We have made careful measurements of the TO phonon dispersion in detwinned NiO at room temperature (RT), and find $E_{\parallel}^{\rm TO}$$-$$E_{\perp}^{\rm TO}$$\sim$$-$1.0 meV in the vicinity of the zone center, in reasonable agreement with the LSDA+$U$ calculations.\cite{J_phonon, Forceconst} In order to understand this, we introduce the anisotropic polarization derived from a Berry's phase. \cite{BerryPol} This anisotropy yields an anisotropic effective Born charge, which was introduced into AFM materials by Ref.~\cite{Massida}, after tremendous success in ferroelectric materials. This spin-dependent term dominates the splitting and provides the missing ingredient to explain the observed lattice distortion and phonon splitting simultaneously.

%Experiment
NiO samples with (111) surface were detwinned in accordance with Ref.~\cite{Slack}, giving the AFM order and lattice contraction along the [111] direction (in this report, indices of cubic symmetry are used). The single structural domain ($T$-domain) was confirmed by the (333) Bragg peak (Fig.~1(a));  a peak from another domain with a different direction for the contraction, if any, would appear at lower angle (the bar in Fig.~1(a)). NiO may also have a different type of domains, originating from  the different spin-axis orientation ($S$-domain). However, Ref.~\cite{Kondo} shows that a field of $>$2 kOe along [1$\bar{1}$0] may be used to align the spin axes along [11$\bar{2}$]. When the field reaches 5 kOe, a single $S$-domain is obtained,\cite{Kondo} but the $T$-domain wall starts to move.\cite{Slack} Considering these facts, magnetic field of 4.5 kOe was applied parallel to [1$\bar{1}$0] during the measurements.

 \begin{figure}
\includegraphics[width=1\linewidth,clip=]{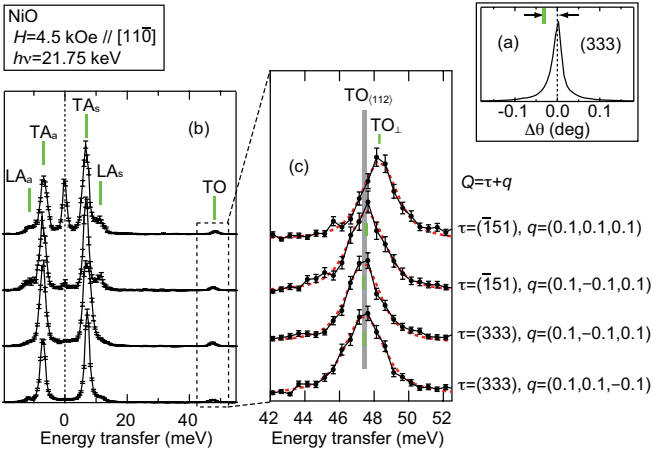}
\caption{\label{fig1} (Color Online) (a) The $\theta$ scan around the (333) Bragg peak. The bar indicates the expected peak position of $\{33\bar{3}\}$. (b) IXS spectra at $|q|$=(0.1, 0.1, 0.1). LA denotes longitudinal acoustic phonon. TA$_{\rm s}$ (LA$_{\rm s}$) indicates the Stokes line of TA (LA), while TA$_{\rm a}$ (LA$_{\rm a}$) does the anti-Stokes line. (c) The enlarged view of (b) at around the TO phonons. Dotted lines are results of curve fitting, and the bars show the peak positions.}
\end{figure}

Phonon measurements using inelastic x-ray scattering (IXS) were performed at BL35XU of SPring-8 \cite{Baron} with a total energy resolution of $\sim$1.6 meV and a momentum resolution of $\sim$0.08 \AA$^{-1} $. The incident energy is 21.75 keV. The measurements were achieved at RT ($\sim$300 K ($<$$T_N$)) to prevent introducing a new domain caused by tiny stresses or thermal gradients.\cite{Slack, Kondo} We also performed similar measurements under 1.7 kOe at some phonon vectors to study field induced effects, such as magnetoelectric effects. The results show the same phonon peak positions as at 4.5 kOe, within an experimental accuracy of $\pm$0.3 meV. We confirmed the zero energy using the midpoint of the Stokes and anti-Stokes lines of the transverse acoustic (TA) phonons. 
 
 Figure 1(b, c) shows the IXS spectra at a phonon vector of $|q|$=(0.1, 0.1, 0.1), where $q$=$Q$$-$$\tau$; $Q$ is the total momentum transfer, and $\tau$ is the nearest Bragg point. A shift is observed, depending on the choice of momentum transfer.  For the top spectrum, where $q$ is parallel to [111],  the polarization vector  of the observed transverse  mode must be in the plane perpendicular to the [111] ordering direction.  For the others,  the situation is less well defined, and we will discuss it again in the context of specific models.  However, naively, one would expect the polarization vectors to be within and perpendicular to the plane defined by the [111] axis and $q$.  In this case, at the chosen total momentum vectors, $Q$, the mode polarized in the plane would tend to dominate the lower three spectra. 

Simulations were carried out  using shell model calculations,\cite{Shellmodel} which describe phonon dispersion, with only a few parameters, very well especially for NiO and MnO.\cite{ MnOShell, Reichardt} The {\sc openphonon} source code\cite{Openphonon} was used for the calculations, and most of the parameters were taken from Ref.~\cite{Reichardt}. \cite{Note} The shell model has several types of parameters including the atomic position, Coulomb interaction (charge), including the core-shell interaction, and short-range  Born-von K\'{a}rm\'{a}n force constants between pairs of atoms. As seen in Fig.~2(a), the calculations assuming the cubic structure reproduce the present experimental phonon dispersion very well. This agreement in dispersion lends credence to the idea that our approach, below, of modifying successively, the various components of the shell model, can be used to understand  the material's behavior.

\begin{figure}
\includegraphics[width=1\linewidth,clip=]{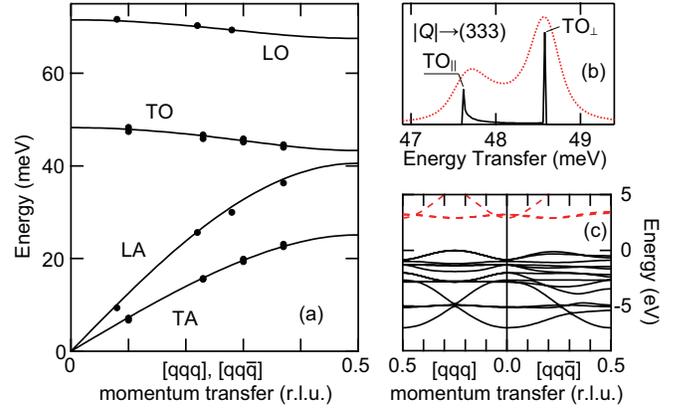}
\caption{\label{fig2}(Color Online) (a) Phonon dispersion of NiO in the $|q|$=[qqq] direction, which is determined theoretically (lines), and experimentally (the dots). (b) Simulated TO phonon distribution in the vicinity of $|Q|$=(333) (solid line). Dotted line shows the spectrum broadened by a 0.4 meV resolution function. (c) The band structure along [qqq] and [qq$\bar{\rm q}$] obtained by LSDA+$U$. Solid lines are the valence bands which are used in calculating the Berry's phase of polarization.}
\end{figure}

First, just the effect of the non-cubic lattice was considered, holding all the other parameters fixed. The experimental lattice distortion of $\sim$90.06$^\circ$ at RT \cite{Slack} gives TO phonon splitting of $\sim$0.06 meV at $|q|$=(0.1, 0.1, 0.1). This value is much smaller than the observed in Fig.~1(c), suggesting that only the lattice distortion can not describe the observed TO peak shift.
 
 Next we introduce additional Born-von K\'{a}rm\'{a}n force constants to simulate the anisotropy of the Heisenberg exchange interaction. Effectively, the shell model includes $J_2$ in the longitudinal force constant between nearest neighbor Ni-O atoms. As for $J_1$, additional force constants along $\langle$01$\overline{1}$$\rangle$ (between parallel spin, $T$) and $\langle$011$\rangle$ (between antiparallel spin, $-$$T$) were included (as the cubic structure assumed, $J_1$ deviation due to the lattice distortion\cite{ExStriction} ($J_1^\pm$) is not considered). The parameter $T$ was  varied to fit the experimental results.

Using these additional force constants, the polarization vectors for the peaks in Fig.~1(c) can be determined as eigenvectors of the dynamical matrix. The spectra at $q$=(0.1, 0.1, $-$0.1) and (0.1, $-$0.1, 0.1) with $\tau$=(333) are assigned to the modes polarized close to [112] (TO$_{[112]}$) and [121] (TO$_{[121]}$) respectively. For $q$=(0.1, $-$0.1, 0.1) with $\tau$=($\bar{1}51$),  the spectrum is dominated by the  TO$_{[121]}$ mode, while the spectra at $q$=(0.1, 0.1, 0.1) with $\tau$=($\bar{1}51$) is assigned to  TO$_{\perp}$. In short, it is estimated from the experiment that the energy of TO$_{\langle112\rangle}$ ($E_{\langle112\rangle}^{\rm TO}$) is $\sim$0.9 meV lower than $E_{\perp}^{\rm TO}$.

The intensity around the zone center is simulated using the parameters  above. Figure 2(b) shows the calculated TO lineshape (solid line) near $|Q|$=(333) consists of two parts; a delta-function-like TO$_{\perp}$ peak and distribution from TO$_{\parallel}$ to TO$_{\perp}$ resulting from the direction dependence of $q $$\to$0. From Fig.~2(b), difference between $E_{\parallel}^{\rm TO}$ and $E_{\perp}^{\rm TO}$ is estimated as $\sim$1.0 meV. The value at RT agrees reasonably with the results of the LSDA+$U$ calculations ($E_{\parallel}^{\rm TO}$$-$$E_{\perp}^{\rm TO}$=$-$1.8 meV),  \cite{J_phonon, Forceconst} which generally assume $T$=0. Furthermore, when this spectrum is convolved with a 0.4 meV resolution function (dotted line), it is clearly seen that TO$_{\perp}$ has stronger intensity than other components. The stronger TO$_{\perp}$ intensity agrees with the observation in MnO. \cite{Chung, MnOIR}

Both the phonon splitting and lattice contraction for NiO are correctly predicted by the LSDA+$U$ calculations.  However, conceptually LSDA+$U$ does not  fit the superexchange mechanism well; the calculated ferromagnetic $J_1$ and lattice contraction require  increasing $|J_1|$ with increasing $r$ in the exchange framework ($r$ indicates that the distance between the nearest-neighbor Ni atoms), since the lattice distortion  is proportional to $-$$\partial$$J_1/\partial$$r$. \cite{ExStriction, Yamada}   Moreover, this contradicts the relation, suggested in Ref.\cite{J_phonon}, between  the sign of $E_{\parallel}^{\rm TO}$$-$$E_{\perp}^{\rm TO}$ and that of $J_1$. Thus some additional ingredient is required,  which is included in the calculations but not in the exchange framework, in order to explain the experimental observation.

The last perturbation we consider is an anisotropic Coulomb interaction. The effective Born charge tensor, $Z^*$, was estimated using LSDA+$U$ as implemented in the {\sc abinit} package.\cite{Abinit} $U$=4 eV was chosen to fit the experimental valence band,\cite{Yin, Shen} mainly because the calculations for the Berry's phase of polarization only require the occupied states (solid lines in Fig.~2(c)). The calculated charge tensor of Ni is
\begin{equation}
Z^*_{\rm Ni}=\left(\begin{array}{ccc}{\rm z}^*& \Delta {\rm z}^* & \Delta {\rm z}^* \\ \Delta {\rm z}^* & {\rm z}^* & \Delta {\rm z}^* \\ \Delta {\rm z}^* & \Delta {\rm z}^* & {\rm z}^*\end{array}\right)
\end{equation}
where z$^*$=2.26 and $ \Delta$z$^*$=0.03. For oxygen ($Z^*_{\rm O}$), z$^*$=$-$2.26 and $ \Delta $z$^*$=$-$0.03 are obtained independently, satisfying the charge neutrality.
This gives an anisotropic parameter, $\Delta $z$^*$/z$^*$=1.3$\times10^{-2}$.
The principal directions and values of the tensor show that each atom has an anisotropic charge of z$^*$+2$\Delta$z$^*$parallel to the AFM order, [111], ($Z^*_{\parallel}$) and z$^*$$-$$\Delta$z$^*$ perpendicular to [111]  ($Z^*_{\perp}$). The anisotropy results from the asymmetric band dispersion; the AFM order brings about band folding in the [qqq] direction (Fig.~2(c)), while no folding occurs in the $\langle$qq$\bar{\rm q}$$\rangle$ directions, leading to the effective charge anisotropy.  Hence this charge anisotropy is spin-dependent, extending, somewhat, the limits of Anderson's original discussion.  \cite{Anderson} 

This anisotropic change may be introduced into the shell model in different ways.  In particular,  the detailed formulation of the shell model  introduces matrices for the total charges $Z$, shell charges $Y$ and core-shell force constants $k$.   We could introduce anisotropy into any of these, $Z$, $Y$, $k$, and here we consider explicitly $Z$ and $Y$.

The free-ion polarizability is given by $Y^2e^2$/$k$ so that we first consider introducing the anisotropy in $Y$.  Assuming the off-diagonal components, $Y_{ij}$ ($i$$\neq$$j$), for both O and Ni have a common value of $Y_{ij}$/$Y_{ii}$=2.8$\times10^{-2}$, we can reproduce the experimental results in Fig 1(c) and all assignments mentioned above.  Based on Ref.\cite{Shellmodel}, which discusses the shell model in detail for a rock-salt structure,  we then estimate the effective  charge anisotropy   as  $\Delta $z$^*$/z$^*$=5.6$\times10^{-2}$ (z$^*$=$\pm$2.21).  

Alternatively, we consider introductions of anisotropy directly in the static charge, $Z$, of the shell model.
Taking $Z_{ij}$/$Z_{ii}$=3$\times10^{-3}$ is sufficient to reproduce the results. This value corresponds to $\Delta $z$^*$/z$^*$=8$\times10^{-3}$ (z$^*$=$\pm$2.19). This agrees better with the LSDA+$U$ calculations than the value from the anisotropic $Y$ case above. However, it is also possible the larger $\Delta $z$^*$/z$^*$ value is caused by lack of consideration of higher order moments than dipole in the shell model.\cite{Shellmodel} 

It is interesting to consider if this anisotropy might be related to the static contraction along the [111] direction. According to Ref.\cite{covalent}, Born effective charges have a correlation to the covalent bonding. This suggests the lattice distortion can be explained using the anisotropic Born charge, as a distance between two atoms at equilibrium may depend on the magnitude of the charges. Noting that the principle directions of $Z^*$ are orthogonal, the lattice distortion from cubic to rhombohedral can be estimated using Ref.\cite{BVS}. In the anisotropic $Z$ case, $\Delta $z$^*$/z$^*$=8$\times10^{-3}$ gives the lattice distortion of 90.15$^\circ$, in fair agreement with the measured 90.06$^\circ$ at room temperature.\cite{Slack}. If we take  $\Delta $z$^*$/z$^*$=5.6$\times10^{-2}$, the distortion in the case of anisotropic $Y$ is estimated, giving a large value, 91.11$^\circ$.

\begin{figure}
\includegraphics[width=1\linewidth,clip=]{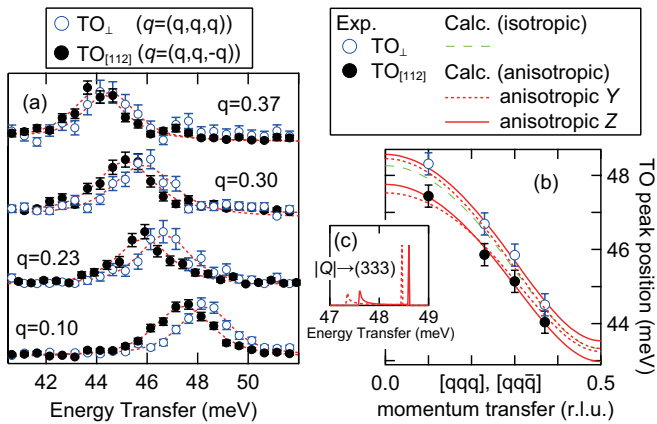}
\caption{\label{fig3} (Color Online) (a) TO phonon dispersion in the [111] and [11$\bar{1}$] directions. Dotted lines are results of curve fitting. (b) Comparison of TO$_{\perp}$ and TO$_{[112]}$ phonon modes between experimental (marks) and theoretical (lines) results. The calculations are based on the shell model with (solid and dotted lines)/without (dashed lines) effective charge anisotropy.  (c) Simulated TO phonon distribution for the effective charge anisotropy in the vicinity of $|Q|$=(333). }
\end{figure}

The measured momentum dependence of the splitting is in agreement with both of these (anisotropic $Y$ or anisotropic $Z$) models as can be seen in Fig.~3 (a,b).  One notes that in all models, the phonon splitting is reduced towards the zone boundary, and the magnitude at the zone boundary depends on the detailed model.  However, our resolution does not allow us to choose between the models. Meanwhile, Figure 3(c) depicts the TO distribution  near $|Q|$=(333) for anisotropic $Y$ and $Z$, showing the similar distribution to Fig.~2(b) for the anisotropic force constant. It should be noted that, based on Ref.\cite{Shellmodel}, it is confirmed that these anisotropic charge models satisfy the generalized Lyddane-Sachs-Teller  relation\cite{CC,PCM} both along and perpendicular to the [111] direction.

In conclusion, we have two different methods of including the anisotropy of the material that fit our experimental results for the phonon splitting: (1) $ad$ $hoc$ addition of force constants to mimic a possible Heisenberg interaction and (2) inclusion of an anisotropy in the electrostatic terms of the shell model.  The latter, which, qualitatively seems to match the anisotropy of the Born effective charge tensor calculated using the LSDA+$U$ method, seems a more reasonable choice.  Furthermore, the latter description provides an additional ingredient that allows one to reconcile the observed lattice contraction with the ferromagnetic $J_1$ interactions calculated using LSDA+$U$.\cite{J_phonon, Forceconst}

These discussions suggest the anisotropic polarization should be  considered  in studying AFM compounds. Both the Berry's phase calculations of polarization and Heisenberg exchange interaction have spin-dependent anisotropy. However, the exchange interaction applies to only magnetic atoms, while the anisotropic effective charge applies to both magnetic and non-magnetic atoms, as required by  charge neutrality. In addition, in order to estimate the exchange interaction precisely, a deviation of the Wannier functions from the cubic symmetry should be considered, even if the lattice is assumed as cubic. This may be relevant to other AFM materials; for example, high-$T_{\rm c}$ cuprates, or iron pnictides.

In summary, we observed TO phonon shifts for detwinned NiO at room temperature, and found that the energy of the TO phonon with atomic motions in the direction of the [111] magnetic ordering is $\sim$1.0 meV lower than the TO mode with motions perpendicular to the ordering direction, near the zone center. We suggest this observation is most simply explained by considering the anisotropic polarization of the medium.  While ultimately related to a spin dependent interaction, at a conceptual level this is an additional ingredient that should be considered, to understand the material behavior, distinct from the exchange interaction. 

The synchrotron radiation experiments were performed at the BL35XU in the SPring-8 with the approval of the Japan Synchrotron Radiation Research Institute (Proposal No. 2008A1584).


\begin{references}
\bibitem[20]{Note}Slight modification of some parameters is achieved to suit the experimental results better; $B'_{12}$=$-$5.07 and $B_{22}$=0.13 ($e^2$/$v$) are different from Ref.~\cite{Reichardt}.
\end{references}
\end{document}